\documentclass[preprint,12pt]{elsarticle}

\usepackage{amssymb}
\usepackage{amsmath} 
\usepackage{amsfonts} 
\def\beq{\begin{equation}}
\def\eeq{\end{equation}}
\def\bea{\begin{eqnarray}}
\def\eea{\end{eqnarray}}

\def\za{\alpha}

\def\ssc{\scriptscriptstyle}

\journal{Annals of Physics}

\begin{document}

\begin{frontmatter}



\title{Ambiguities and Subtleties in Fermion Mass Terms in Practical Quantum Field Theory}
\author{Yifan Cheng}
\ead{s1222027@cc.ncu.edu.tw}

\author{Otto C. W. Kong\corref{cor1}}
\ead{otto@phy.ncu.edu.tw}
\cortext[cor1]{Corresponding author. Tel.: +886 3 4227151 ext. 65372.}

\address{Department of Physics and 
Center for Mathematics and Theoretical Physics,\\
National Central University, Chung-Li 32054, Taiwan}

\begin{abstract}
This is a review on structure of the fermion mass terms 
in quantum field theory, under the perspective of
its practical applications in the real physics of Nature ---
specifically, we discuss fermion mass structure in the Standard
Model of high energy physics, which successfully describes fundamental
physics up to the TeV scale.
The review is meant to be pedagogical, with
detailed mathematics presented beyond the level one can find 
any easily in the textbooks. The discussions, however, bring 
up important subtleties and ambiguities about the subject that
may be less than well appreciated. In fact, the naive perspective
of the nature and masses of fermions as one would easily drawn
from the presentations of fermion fields and their equations
of motion from a typical textbook on quantum field theory leads
to some confusing or even wrong statements which we clarify
here. In particular, we illustrate clearly that a Dirac fermion
mass eigenstate is mathematically equivalent to two degenerated
Majorana fermion mass eigenstates at least so long as the mass
terms are concerned. There are further ambiguities and subtleties 
in the exact description of the eigenstate(s). Especially, for the 
case of neutrinos, the use of the Dirac or Majorana 
terminology may be mostly a matter of choice. The common usage of 
such terminology is rather based on the broken $SU(2)$ charges
of the related Weyl spinors hence conventional and may not
be unambiguously extended to cover more complicate models.   
\end{abstract}

\begin{keyword}
Dirac Mass \sep Majorana Mass
\PACS 11.10.-z \sep 14.60.Pq
\end{keyword}
\end{frontmatter}


\section{Introduction}
This article has its origin from almost twenty years ago when one
of us (O.K.) was very interested in fermion/neutrino mass modeling 
and found the common question of if neutrinos are Majorana or Dirac
particles quite a confusing one. The author was contemplating a
`Majorana' mass matrix for the then plausible maximal atmospheric 
neutrino mixing which put the muon and tau neutrinos ($\nu_\mu$
and $\nu_\tau$) as essentially a `Dirac' pair, with one taken as
the antiparticle of the other. We were more recently
drawn to clarify further some issues about the meaning of the
Dirac versus Majorana question, which is conceptually simply a fermion
mass structure question. We consider fully clarifying the question
in an article useful to the community, at least in the pedagogical 
sense. That is what we set out to do here. We will draw explicit
examples for the fermion mass terms from 
quantum field theory, with more focus on the
neutrinos.

Another uncomfortable question we sometimes have to confront during a 
presentation is the question `why the mass matrix is not symmetric 
(hermitian)'. In a quantum field theory textbook, both the Dirac and
Majorana mass terms are hermitian, hence a mass matrix is always a hermitian 
one. However, that does not quite work for fermion mass matrices 
in the Standard Model (SM) of high energy physics. The 
Weyl fermion field or two component spinor, if discussed in such textbooks, 
is said to be massless by definition. In practice, only (Weyl) two-spinors 
are to be used in the SM, the massless fermionic fields of which of course 
get mass after electroweak symmetry breaking. Of course two-spinors can be
used to write the mass terms, and mass eigenstate two-spinors can be
found. Casting the latter as Majorana or Dirac four-spinors, however, 
may not be always trivial. 
Phenomenologists working on SM physics and beyond 
talk about fermions being Dirac or Majorana in a specific way not
in direct correspondence with the naive description in standard quantum
field theory, where no reference is made to properties of the component
two-spinors under the broken symmetries.
 There are quite some ambiguities and 
subtleties behind the story of fermion masses, some aspects of which 
even some experts in the field may not fully realize. 

A key feature of the SM is that the fermionic
field content consists of families of 15 
spinor fields, in 5 different gauge multiplets, which are all 
chiral. In fact, each family is a minimal chiral set with nontrivial 
cancellation  of all gauge anomalies \cite{GM,O}. As such, the structure 
has completely no parity symmetry. The Lee-Yang discovery of parity 
violation, in our opinion, should have put the prejudice of parity being a
part of the fundamental (spacetime) symmetry to an end. The irreducible 
spinor representation of spacetime symmetry then is given by the chiral, 
two component, spinor, not the four-spinors as in Dirac or Majorana. So, 
the chiral spinor, or called Weyl spinor, is the base we use here to 
clarify all the issues. And we will do it in full detail. From the 
perspective of anomaly cancellation, as well as a supersymmetry
generalization of the SM, it is more natural to consider all the 
3 $\times$ 15 spinor states as of the same handedness. The basic $SU(2)_L$ 
singlet states, for instance, are to be considered as left-handed
antiquark and positively charged lepton states.

\section{Dirac Mass and Dirac Fermion}
We first recall the description of Dirac fermion and its mass term
more or less as usually done in textbooks, with focus however on 
the fundamental description on them in terms of two-component 
left-handed spinors, under the perspective of SM fermions
described above. Before electroweak symmetry breaking, all fermionic 
states in the SM are indeed chiral Weyl spinor and massless as
the term usually requires. Mass terms are not allowed
by the gauge symmetries, as no two of the spinor multiplets match
as a pair of conjugate quantum numbers. The latter is the true meaning
of the fermion field spectrum being chiral, no vectorlike pairs. 
In our subsequent discussion, the term
spinor when not specified always means a two-spinor, mostly taken
as left-handed. 

A Dirac mass term between two (left-handed) spinors
$\psi$ and $\chi$ can be written as 
\bea
 i   m\chi^{\ssc T} \bar{\sigma}^{\ssc 2} \psi
- i   m^*\psi^{\dagger}  \bar{\sigma}^{\ssc 2} \chi^* 
\;, \label{mD}
\eea
where we have contrived to use a complex mass parameter $m$.\footnote{
Our basic notation follows mostly that of the text by Maggiore \cite{Maggiore}.} The idea is
to keep everything in the generic admissible setting, which helps better to
illustrate any subtle difference among things. With no other mass term 
involving the pair of spinors $\psi$ and $\chi$, one can absorb any 
complex phase in $m$, say into $\psi$. The spinors can be 
called a Dirac pair and put together as parts of one Dirac four-spinor 
\beq \label{Ds}
\Psi = \left( \begin{array}{c} 
 \psi \\
- i \bar{\sigma}^2\chi^*
\end{array} \right) \;,
\eeq 
describing a (mass eigenstate) Dirac fermion with a corresponding Dirac 
particle picture. The familiar form of Dirac mass term can be 
expanded, in the chiral representation which we adhere to, as
\beq
  m_{\!\ssc D} \overline{\Psi}\Psi 
=
\left( \begin{array}{cc} \psi^{\dagger} 
& i\chi^{\ssc T} \!\bar{\sigma}^{\ssc 2} \end{array}\right)
\left( \begin{array}{cc} 
 0 &    m_{\!\ssc D} \\
    m_{\!\ssc D} & 0   
\end{array} \right)
\left( \begin{array}{c} 
 \psi \\
- i\bar{\sigma}^{\ssc 2} \chi^*
\end{array} \right)   \;,
\eeq
which gives the above expression for $m=m^*=m_{\!\ssc D}$ real and positive.
Note that $\Psi$ is composed of a left-handed part $\psi$ and a right-handed 
part $\chi^{\ssc C} = - i \bar{\sigma}^{\ssc 2}\chi^*$ which is a right-handed 
spinor conjugate to the other left-handed spinor $\chi$ independent of $\psi$.
In other words, $\psi$ and $\chi$ make a vectorlike pair.

For the readers to whom the above expressions may appear somewhat uncomfortable,
we offer some explanation here. Firstly, $- i \bar{\sigma}^{\ssc 2}$, which 
equals to the more commonly used $i {\sigma}^{\ssc 2}$ as a matrix, 
is used here to give the charge conjugate $\chi^{\ssc C}$ of $\chi$. 
It is more proper. In a bit more detail, we have spin components
$\psi_{\ssc\za}$ and $\chi_{\ssc\za}$, and 
 $\chi^{\ssc C \dot{\za}}\equiv \bar{\chi}^{\ssc\dot{\za}}$ and
$\bar{\sigma}^{\ssc 2}$ with components $\bar{\sigma}^{\ssc 2\dot{\za}\za}$ giving 
$\bar{\chi}^{\ssc\dot{\za}} = -i \bar{\sigma}^{\ssc 2\dot{\za}\za} \chi_{\ssc\za}^*$.
See for example appendix A in the classic text on supersymmetry from Wess
and Bagger \cite{WB}, to the notation of which we followed. However, to simplify
expressions, spin indices are mostly suppressed. Left-handed
spinors like $\psi$ when unspecified carries lower spin index $\za$ while 
right-handed spinors like $\bar{\psi}$ or $\psi^{\ssc C}$ carries upper
spin index $\dot{\za}$. Note that $\chi^{\ssc T}$ and $\psi^\dag$, in expression 
(\ref{mD}), as spinors carry lower $\dot{\za}$ index. In fact, the language 
of supersymmetric field theory offers a viewpoint that helps to clarify many 
issues involved here and we will be exploiting that in our discussions. A 
(chiral) fermion as a (two-)spinor is part of a chiral superfield, which is a 
scalar field on the (chiral) superspace. As such, the spinor is left-handed. 
Fermion mass terms are to be written as bilinear terms in chiral superfields 
within the holomorphic superpotential, hence bilinear in left-handed
spinors. The fermion mass term resulted has the form 
$m \chi \cdot \psi \equiv m \chi^{\ssc\za}  \psi_{\ssc\za} = m \psi \cdot \chi$~,
symmetric in the spinors or the two superfields. The term is exactly the same
as the first term in expression (\ref{mD}). The conjugate term needed to 
maintain the hermitian nature of the Lagrangian density is given by the
anti-holomorphic part  $m^* \bar{\psi} \cdot \bar{\chi} 
= m^* \bar{\psi}_{\ssc\dot{\za}} \bar{\chi}^{\ssc\dot{\za}}$~, 
which gives the second term. If that is the only mass 
term(s) involving the two spinors(/superfields), the complex phase can be 
absorbed through (super)field re-definition. We will see below that the 
Dirac mass $m_{\!\ssc D}$ is really a mass eigenvalue, and as such real and 
positive. Note that we cannot write the two terms in expression (\ref{mD}) as 
the single Dirac four-spinor mass term without first removing the complex phase 
in $m$. For the readers who do not believe supersymmetry have anything 
to do nature, we note that one can add heavy enough soft supersymmetry breaking 
masses to decouple all the scalar superparticles of the chiral superfields from 
a (hypothetical) supersymmetric model to retrieve model with only the 
spinor parts. We use it here only as a theoretical tool to illustrate
the less familiar structure in theory with two-spinors, which is difficult
to be found given in much field theoretical detail as we do here. 

In the case of the SM, we have for each of the sectors of charged lepton, 
up-type quarks, and down-type quarks six (for the quark cases colored) chiral 
spinors three from $SU(2)_L$ doublets and three from
singlets. The parts from the singlets have conjugate charge and color from
the corresponding parts from the doublets. In the generic flavor basis,
each sector has nine Dirac mass terms of the form given in expression (\ref{mD})
in which $\psi$ can be considered one from a doublet and $\chi$ one from 
a singlet. All nine different complex $m$ values, say in the
$i  m \chi^{\ssc T} \bar{\sigma}^2 \psi$
part can be taken together as a `Dirac mass matrix'. The diagonalization of 
the latter requires a bi-unitary transformation and gives the mass eigenstates 
with three matching pairs of $\psi$ and $\chi$. The (nine) two-spinor Dirac 
mass terms actually cannot be written in the form  
$m_{\!\ssc D} \overline{\Psi}\Psi$ before diagonalizing the full
mass matrix to find out which $\psi$ pairs with which $\chi$. We cannot get
rid of all the complex phases in general, which is related to the source of 
the Kobayashi-Maskawa 
phase for CP violation in quark physics for instances. More importantly, 
without knowledge of the right pairing, there are in the case at hand nine 
possible $\Psi$'s we can form. So, the `Dirac mass matrix' is NOT a mass 
matrix for a set of (three) Dirac four-spinors to be diagonalized to obtain the 
mass eigenstates. Hence, straightly speaking, the term Dirac fermion can be 
applied really only to the mass eigenstates, with real and positive eigenvalues. 
Note that while the fact that the two component two-spinors of 
each Dirac spinor have conjugate charge and color may be considered a 
required feature in the quantum field theory definition of the Dirac field,
their coming from a doublet and a singlet of $SU(2)_L$ is not, and should not.

For the neutrinos, the singlet fields are not available in the SM 
fermionic family spectrum as described. They are frequently added in 
neutrino mass models, as so-called `right-handed neutrinos' or `sterile 
neutrinos'. Described as left-handed two-spinors with their antiparticles, 
the latter two have no difference. We use here  
only the term singlet neutrinos. Recall that we take 
only left-handed two-spinors as the basic input ingredient in our discussions. 
We can take a SM neutrino state, {\em i.e.} one from a $SU(2)_L$ doublet,
as $\psi$ and a (left-handed) singlet neutrino $\nu_{\!\ssc S}$ as $\chi$ 
with a Dirac mass term as in expression (\ref{mD}). With three singlet 
neutrinos, we can have a `Dirac mass matrix'. All that is exactly analogous 
to the quarks and charged leptons. 
Phenomenologists and experimentalists typically reserve the
term Dirac neutrinos only for the kind of massive neutrinos.
However, the story of the neutrino
mass can be more complicated, namely Majorana mass terms are also admissible.
The latter are a kind of self mass term in the language of two-spinors. 
Moreover, we can take a Dirac like mass term with both $\psi$ and $\chi$
as the SM neutrinos, like $\nu_\tau$ and $\nu_\mu$. Such a mass term
is however usually called a Majorana mass term instead.  The even more 
provoking statement is : in the case that such a Dirac mass term given as
the two-spinor mass terms in the form of expression (\ref{mD})  are the 
only mass terms involving $\nu_\tau$ and $\nu_\mu$, the two form a sort 
of Dirac spinor with four complex spin degrees of freedom! We will illustrate
the story behind the potentially confusing statement below. For those who
may object to call the $\nu_\mu$-$\nu_\tau$ mass term Dirac, we have to
say that they have a point, but would have to define the Dirac mass term with
clear criterion on the $SU(2)_{L}$ nature of the two-spinors involved
as well as their role in the mass eigenstate to justify the restricted
usage. Without the extra elements in the definition of a Dirac mass term 
or Dirac fermion based on the two-spinors, the objection cannot be justified.
We will get back to the question in Sec.4 and 5 below.

\section{Majorana Mass and Majorana Fermion}
A Majorana mass term for a two-spinor $\psi$ is a self mass
term in the form
\beq \label{mM}
\frac{1}{2} m_{\!\ssc M} \overline{\Psi}_{\!\ssc M} \Psi_{\!\ssc M} 
= \frac{1}{2} m_{\!\ssc M}   (\psi^{\ssc C})^\dag   \psi   
+  \frac{1}{2} m_{\!\ssc M}\psi^{\dag}\psi^{\ssc C} 
\eeq
where the Majorana four-spinor $\Psi_{\!M}$ is given by
\beq \label{Ms}
\Psi_{\!\ssc M} = \left( \begin{array}{c} 
\psi \\ 
- i\bar{\sigma}^{\ssc 2} \psi^*  
\end{array} \right) 
= \left( \begin{array}{c} 
\psi \\ 
 \psi^{\ssc C} 
\end{array} \right)\;,
\eeq
with $m_{\!\ssc M}$ being real and positive.
It does have four real (spin) degrees of freedom as in the Majorana
representation given here as the two complex spin components of $\psi$, 
and $\Psi_{\!\ssc M}^{\ssc C}=\Psi_{\!\ssc M}$. 
The term violates any $U(1)$ symmetry carried by $\psi$, as
 $(\psi^{\ssc C})^\dag$ carries the same charge as $\psi$. The 
electromagnetic symmetry and the lepton number symmetry are the usually 
considered gauge and global versions, respectively, of such a $U(1)$
symmetry. Hence, neutrinos are the only SM fermions that can have Majorana 
masses, which then imply lepton number violation. If the above term
in Eq.(\ref{mM}) is the only mass term involving $\psi$, we can call 
it a Majorana fermion. 

We can write mass terms for two Majorana four-spinors 
$\Psi_{\!\ssc M}$ and $\Phi_{\!\ssc M}= ( \begin{array}{cc} 
\chi & \chi^{\!\ssc C}
\end{array} )^{\!\ssc T}$
as
\bea
- {\mathcal L}_m^{\ssc (4)} 
&=& \frac{1}{2} \left( \begin{array}{cc} 
\overline{\Psi}_{\!\ssc M} & \overline{\Phi}_{\!\ssc M}
\end{array} \right) 
{\mathcal M}_{\!\ssc R} 
\left( \begin{array}{c} 
{\Psi}_{\!\ssc M}  \\ {\Phi}_{\!\ssc M}
\end{array} \right)
\nonumber \\
&=&
\frac{1}{2} m_{\!\ssc 1} \overline{\Psi}_{\!\ssc M} \Psi_{\!\ssc M} 
+ \frac{1}{2} m_{\!\ssc 2} \overline{\Phi}_{\!\ssc M} \Phi_{\!\ssc M} 
+ \frac{1}{2} m_{\!\ssc 1\!2} \overline{\Psi}_{\!\ssc M} \Phi_{\!\ssc M} 
+ \frac{1}{2} m_{\!\ssc 2\!1} \overline{\Phi}_{\!\ssc M} \Psi_{\!\ssc M} \;,
\label{(4)}
\eea
where we take all mass parameters as real and 
$m_{\!\ssc 1\!2}= m_{\!\ssc 2\!1}$.  The mass
eigenstate Majorana fermions are to be obtained through diagonalizing
the real and symmetric matrix ${\mathcal M}_{\!\ssc R}$. For the case 
$m_{\!\ssc 1}=m_{\!\ssc 2}=0$, we have maximal mixing with the mass
matrix to be diagonalized by a simple $\frac{\pi}{4}$ rotation to give
mass eigenstates
\bea
 \frac{1}{\sqrt{2}}
\left(  \Psi_{\!\ssc M}  +  \Phi_{\!\ssc M} 
 \right)
\qquad\qquad  {\mbox{and}} \qquad\qquad
\frac{1}{\sqrt{2}}
\left(  \Psi_{\!\ssc M}  - \Phi_{\!\ssc M} 
 \right)\;,
\eea
with corresponding mass eigenvalues $m_{\!\ssc 1\!2}$ and $-m_{\!\ssc 1\!2}$. 
The Majorana four-spinor mass eigenstates
imply the two-spinor mass eigenstates
\bea
 \frac{1}{\sqrt{2}}
\left(  \psi +  \chi 
 \right)
\qquad\qquad  {\mbox{and}} \qquad\qquad  
\frac{1}{\sqrt{2}}
\left(   \psi -  \chi 
 \right)\;. \label{esM}
\eea
All that appears quite trivial so far, except that we have one negative
mass eigenvalue the sign of which cannot be fixed by any unitary 
transformation, that cannot change the zero trace of the mass matrix.

\section{Dirac or Majorana fermions as two-spinor mass eigenstates}
Let us look at the `Majorana mass matrix' for the above case
using only two-spinors. The mass terms in Eq.(\ref{(4)}) can be 
written equivalently as 
\bea
- {\mathcal L}_m^{\ssc (2)}
&=& \frac{1}{2} \left[
m_{\!\ssc 1}   (\psi^{\ssc C})^\dag   \psi 
+ m_{\!\ssc 2}   (\chi^{\ssc C})^\dag   \chi
+  m_{\!\ssc 1\!2} (\psi^{\ssc C})^\dag  \chi
+  m_{\!\ssc 2\!1} (\chi^{\ssc C})^\dag   \psi \right]
\nonumber \\
&& + \frac{1}{2} \left[
m_{\!\ssc 1}   \psi^\dag \psi^{\ssc C}
+ m_{\!\ssc 2}     \chi^\dag \chi^{\ssc C}
+  m_{\!\ssc 1\!2}  \psi^\dag  \chi^{\ssc C}  
+  m_{\!\ssc 2\!1}  \chi^\dag \psi^{\ssc C}   \right]
\nonumber \\
&=& \frac{1}{2} \left( \begin{array}{cc} 
(\psi^{\ssc C})^\dag   &  (\chi^{\ssc C})^\dag 
\end{array} \right) 
{\mathcal M}_{\!\ssc R}  
\left( \begin{array}{c} 
\psi  \\ \chi
\end{array} \right)
+ \frac{1}{2} \left( \begin{array}{cc} 
\psi^\dag  &  \chi^\dag
\end{array} \right) 
{\mathcal M}_{\!\ssc R}^\dag 
\left( \begin{array}{c} 
\psi^{\ssc C}  \\ \chi^{\ssc C}
\end{array} \right)
 \;,  \label{Mm2}
\eea
where we have contrived to write ${\mathcal M}_{\!\ssc R}^\dag$ in the
place of the equivalent  ${\mathcal M}_{\!\ssc R}$ in the last term.
The mass matrix may hence also be taken as one for the two-spinors. 
Let us again use the superfield language to help clarify things
further. A chiral superfield self mass term gives fermion mass
in the form 
$\frac{1}{2}m \psi \cdot \psi = \frac{1}{2}m \psi^{\ssc\za} \psi_{\ssc\za}$ 
which is the same as $\frac{i}{2} m\psi^{\ssc T} \bar{\sigma}^{\ssc 2} \psi
= \frac{1}{2} m\, (\psi^{\ssc C})^\dag \psi$, except that $m$ is generally
complex. The conjugate gives 
$\frac{1}{2} m^* \bar{\psi} \cdot \bar{\psi}=
-\frac{i}{2} m^* \psi^\dag \bar{\sigma}^{\ssc 2} \psi^*
=\frac{1}{2} m^* \psi^\dag \psi^{\ssc C}$. Like the case for the Dirac
four-spinor, we have to absorb the complex phase in $m$, as going to mass 
eigenstate to write the Majorana four-spinor. For the generic case with a
number of (two-)spinors, the mass matrix being only the first half of 
the last line in Eq.(\ref{Mm2}) is not hermitian. Recall 
that $\chi \cdot \psi=\psi \cdot \chi$; superfields are superspace 
scalars and their mass matrix symmetric. With two superfields the 
mass terms are to be written in the form of Eq.(\ref{Mm2}) with however
complex mass parameters. The mass matrix ${\mathcal M}$ is then complex
symmetric, hence not hermitian;  
${\mathcal M}^\dag= {\mathcal M}^* \ne {\mathcal M}$,
which is why we write ${\mathcal M}_{\!\ssc R}^\dag$ in the above 
equation though ${\mathcal M}_{\!\ssc R}^\dag={\mathcal M}_{\!\ssc R}$.
The case for ${\mathcal M}$ with $m_{\!\ssc 1}=m_{\!\ssc 2}=0$ is, however,
particularly intriguing. On one hand, it matches with the results
discussed above in the previous section for the Majorana four-spinor picture. 
On the other hand, it fits the description of a single Dirac fermion. The 
latter is illustrated explicitly by 
\bea
\frac{1}{2}  m\, (\psi^{\ssc C})^\dag  \chi
+ \frac{1}{2}  m\, (\chi^{\ssc C})^\dag   \psi 
= m \chi \cdot \psi 
&=& i m \chi^{\ssc T} \bar{\sigma}^{\ssc 2} \psi  \;,
\nonumber \\
\frac{1}{2}  m^* \psi^\dag  \chi^{\ssc C}
+ \frac{1}{2}  m^* \chi^\dag   \psi^{\ssc C}
=  m^* \bar{\psi} \cdot \bar{\chi}
&=& -i m^* \psi^\dag \bar{\sigma}^{\ssc 2} \chi^* \;.
\eea 
For readers familiar with the superfield language, this would be 
no surprise at all. Mass term with two different superfields does 
not distinguish between Majorana or Dirac. It is one and gives one 
fermion mass term in terms of two-spinors. One can diagonalize the
mass matrix ${\mathcal M}$ for the case as  
\begin{equation} \label{dia}
U^{\ssc T}
\left( \begin{array}{cc} 
 0 & m\\
 m & 0
\end{array} \right)
U
= \left( \begin{array}{cc} 
  \left|m\right|  & 0 \\
 0  & \left|m\right|
\end{array} \right) \;,
\end{equation}
where $m=\left|m\right|e^{\ssc i\theta}$ and 
\beq \label{u}
U=\left( \begin{array}{cc} 
 \cos\frac{\pi}{4}  & -i\sin\frac{\pi}{4} \\
 \sin\frac{\pi}{4} & i\cos\frac{\pi}{4}
 \end{array} \right) 
e^{\ssc \frac{-i\theta}{2}} \;,
\eeq
giving mass eigenstates 
\bea
 \frac{1}{\sqrt{2}} \, e^{\ssc\frac{i\theta}{2}}
\left(  \psi +  \chi 
 \right)
\qquad\qquad  {\mbox{and}} \qquad\qquad  
i \frac{1}{\sqrt{2}} \, e^{\ssc\frac{i\theta}{2}}
\left(   \psi -  \chi 
 \right)\;. \label{es}
\eea
The hermitian conjugate part,   $m^* \bar{\psi} \cdot \bar{\chi}$~,
with mass matrix ${\mathcal M}^\dag={\mathcal M}^*$ is to be 
diagonalized similarly with the unitary matrix $U^*$.
The corresponding mass eigenstates are then given as
\bea
 \frac{1}{\sqrt{2}} \, e^{\ssc\frac{-i\theta}{2}}
\left(  \psi^{\ssc C} +  \chi^{\ssc C} 
 \right)
\qquad\qquad  {\mbox{and}} \qquad\qquad  
-i \frac{1}{\sqrt{2}} \, e^{\ssc\frac{-i\theta}{2}}
\left(   \psi^{\ssc C} -  \chi^{\ssc C} 
 \right)\;, \label{es.}
\eea
which are indeed the charge conjugate to eigenstates right above. 
Comparing eigenstates in expression (\ref{es}) versus those in 
expression (\ref{esM}) obtained from working on the Majorana four-spinors
illustrates something interesting. The only difference is phase factors,
which are physically insignificant. In fact, one should put $\theta=0$ 
for a meaningful comparison, which then yields a factor of $i$
difference only in the expression of the second eigenstates. Arbitrary
phase factors may actually be put into each of the eigenstates in 
(\ref{esM}). To give the mass eigenvalues for ${\mathcal M}$ as real 
and positive, however, requires fixed specific phases in the expressions 
of the eigenstate in (\ref{es}).  In fact, any complex symmetric matrix 
for the two-spinor mass terms can be properly diagonalized to give real 
and positive eigenvalues. The diagonalization by a unitary rotation of 
the states is of course not the usual unitary transformation for the 
mass matrix [{\em cf.} Eq.(\ref{dia})]. As such, one cannot take 
arbitrary linear combinations of the two degenerated mass eigenstates 
as an eigenstate. Mathematically, even changing an eigenstate by a 
phase factor changes the corresponding mass eigenvalue (by a phase 
factor which has no physical meaning). The negative sign in one of the 
mass eigenvalues as given in the analysis using Majorana four-spinors
is now removed, showing the states as actually degenerate in mass. 

At this point, it is also interesting to comment on the issue of
Majorana phases in neutrino mass matrix. First of all, if one looks at
a naive Majorana mass matrix defined as one for the mass terms of Majorana 
four-spinors like the one in Eq.(\ref{(4)}), there would not be any 
Majorana phases as all entries are by definition real. For the general 
mass matrix ${\cal M}$ discussed here, as given in terms of two-spinors, 
there can be complex matrix elements and hence Majorana phases, which 
provide a source of CP-violation beyond that of the Dirac phase. 
The latter is a relative phase between the mass matrices of the SM 
neutrinos and charged leptons, as in the case of Kobayashi-Maskawa 
phase between the quark sectors. The diagonalizing unitary matrix 
can be written in two parts as $U=RP$, where the real rotation $R$
gives $R^{\ssc T}{\cal M}R$ as a diagonal matrix of complex 
`eigenvalues' while $P$ is a diagonal matrix of the complex
phases.  Only the relative phases in $P$ are physical. They are the 
Majorana phases \cite{Mohapatra}. For instance, for $U$ in 
Eq.(\ref{dia}), we have $R$ as a 45$^o$ rotation and $P$ given as
$\rm{diag}\{e^{\ssc\frac{-i\theta}{2}},e^{\ssc \frac{i\left(\pi-\theta\right)}{2}}\}$, 
showing a Majorana phase of $e^{i\frac{\pi}{2}}$. The latter  is to 
be matched with the relative $i$ factor in the eigenstates as given
by Eq.(\ref{es}); the common $e^{\ssc\frac{i\theta}{2}}$ phase is 
not physically observable.

To summarize the above:
a generic complex `Majorana mass matrix' of the kind ${\mathcal M}$,  
involving any number of (two-)spinors is to be diagonalized with mass 
eigenstate spinors each of which can be used to write a Majorana 
four-spinor as in Eq.(\ref{Ms}). The complex matrix itself cannot, 
however, be considered as a mass matrix for Majorana four-spinors. 
In the most general case, one will be dealing with an $n\times n$ 
matrix with $n$ possibly odd and a number of physical complex phases 
contained in it. Any eigenstate spinor $\psi$ may then be written 
as a Majorana four-spinors as in Eq.(\ref{Ms}). The treatment 
using two-spinors instead of Majorana four-spinors is hence clearly
a better way. When there are pairs of degenerate eigenstates, 
one can invert the unitary transformation given by matrix $U$ in 
Eq.(\ref{u}) to obtain a $\psi$-$\chi$ pair and rewrite the mass 
terms for the two spinors as a single Dirac mass term using the Dirac 
four-spinor given as in Eq.(\ref{Ds}). In this sense, so long as the 
mathematical content of the mass term(s) is concerned, a Dirac fermion 
is the same as two mass degenerate two-spinor or Majorana fermions. 
Explicitly, starting with two Majorana four-spinors with degenerate
mass $m_r$ (real and positive), we have 
\bea
&& \frac{1}{2} m_r \overline{\Psi}_{\!\ssc M}' \Psi_{\!\ssc M}' 
+\frac{1}{2} m_r \overline{\Phi}_{\!\ssc M}' \Phi_{\!\ssc M}' 
= \frac{1}{2} m_r   (\psi^{'\ssc C})^\dag   \psi' 
+ \frac{1}{2} m_r  (\chi^{'\ssc C})^\dag   \chi'
+ \rm{h.c.}
\nonumber \\
 = && \frac{1}{2}  m_r \left[
 (\psi^{\ssc C})^\dag  \chi +
 (\chi^{\ssc C})^\dag   \psi \right] + \rm{h.c.}
= m_r \overline{\Psi}_{\!\ssc D}{\Psi}_{\!\ssc D} \;,
\label{key}
\eea
where the Dirac four-spinor 
${\Psi}_{\!\ssc D}\equiv \left( \begin{array}{cc} \psi  & \chi^{\ssc C}
\end{array} \right)^{\!\ssc T}$ with $\psi=\frac{1}{\sqrt{2}} (\psi' -i\chi')$
and  $\chi=\frac{1}{\sqrt{2}} (\psi' +i\chi')$.

Some reader may jump onto the last statement and disagree, saying 
that a Dirac fermion has the two-spinor pairs $\psi$ and $\chi$ 
having opposite charges (conjugate quantum numbers) while Majorana 
fermions are self-conjugates. We certainly have no objection to 
use only the term Dirac for fermions like mass eigenstates of quarks 
and charged leptons. We only say that the structure of the mass terms 
are blind to the quantum numbers of the two-spinors involved and 
hence do not distinguish a Dirac fermionic state from two degenerate 
Majorana fermionic states. Hence, when the states $\psi$ and $\chi$
have no conserved quantum numbers, like the case of the neutrinos,
the question about if the particles are Dirac or Majorana could
be really misleading. Let us elaborate further the advantage of
the Dirac terminology through exploring the Dirac mass 
term. Consider the `mass matrix' as in
\beq
 m \,(\chi^{\ssc C})^\dag   \psi +m^* \psi^\dag  \chi^{\ssc C}
=\left( \begin{array}{cc} 
  \psi^\dag & (\chi^{\ssc C})^\dag 
\end{array} \right)
\left( \begin{array}{cc} 
 0 & m^*\\
 m & 0
\end{array} \right)
\left( \begin{array}{c} 
  \psi   \\   \chi^{\ssc C}
\end{array} \right)  \;,
\eeq
which also gives the full `Dirac' mass term the same as the
`Majorana' mass terms in Eq.(\ref{Mm2}) for ${\mathcal M}$ with 
$m_{\!\ssc 1}=m_{\!\ssc 2}=0$. It is important to note that we 
here take a single hermitian Dirac mass term in a matrix form,
not only the left-handed part that split into two terms as in the above
analysis. It yields mass eigenstates  
\bea
 \frac{1}{\sqrt{2}} \left(
e^{\ssc{i\theta}} \psi + \chi^{\ssc C}  
 \right)
\qquad\qquad  {\mbox{and}} \qquad\qquad  
\frac{1}{\sqrt{2}} \left(
e^{\ssc{i\theta}} \psi - \chi^{\ssc C}  
 \right)\;, \label{esD}
\eea
with corresponding eigenvalues $|m|$ and $-|m|$, respectively. 
Again with the zero trace, the negative sign in one of the eigenvalues 
cannot be removed. The mass eigenstates look very different from those 
in expression (\ref{es}). The latter involves combinations of $\psi$
and $\chi$, while the eigenstates here involve combinations of $\psi$
and $\chi^{\ssc C}$. In the case of the quarks and the charged leptons,
each pair of $\psi$ and $\chi$ has conjugate conserved quantum numbers,
color and electric charge. The formal eigenstates as combinations of 
$\psi$ and $\chi$ violate them hence are not good to use. Note though 
the mass terms preserve the symmetries. Combinations of $\psi$ and 
$\chi^{\ssc C}$, however, mix only handedness and quantum number of 
the broken $SU(2)_{L}$. That speaks further in favor of the 
Dirac mass terminology for the cases. Actually, the mass term from 
each individual mass eigenstate of (\ref{esD}), {\em i.e.}
$\frac{\pm|m|}{2} \left(e^{\ssc{i\theta}} \psi \pm \chi^{\ssc C} \right)^\dag
\left(e^{\ssc{i\theta}} \psi \pm \chi^{\ssc C} \right)$,  contains 
terms of $\psi^\dag \psi$ and $(\chi^{\ssc C})^\dag \chi^{\ssc C}$ 
only to be canceled between the two eigenstates. Similar feature 
goes with the eigenstates of (\ref{es}), there with terms of 
$(\psi^{\ssc C})^\dag  \psi$ and $(\chi^{\ssc C})^\dag \chi$.
None of all that shows up in the expansion of the mass terms 
written in terms of Dirac or Majorana four-spinors. The four-spinor 
expressions have clear advantage once we get to the mass eigenstates.
Conventional quantum field theoretical calculations are also 
performed with four-spinors.

The case for the neutrinos is more tricky.
One may want to say that the Dirac term should
be used when the states $\psi$ and $\chi$ have different quantum
numbers to begin with, no matter those quantum numbers are
conserved or not. Such quantum numbers are certainly meaningful
if they were connected to broken gauge symmetries. In particular,
one may want to call a neutrino mass eigenstate Dirac if the
two-spinors involved one from a $SU(2)_{L}$ doublet while
the other from a singlet, like the quarks and the charged leptons,
and call it Majorana if it involved states from only doublets
or singlets. In fact, this is the more common usage.
But the Dirac terminology can only be used when
the pair of two-spinor mass eigenstates are exact equal mixtures of
the two (doublet and singlet) parts. Moreover, the usage requires 
`defining' what is a Dirac fermion with criteria beyond the nature 
of the free field.

\section{Three-family Majorana and Dirac Mass Matrices}
To complete the story, we address explicitly the full fermion mass 
matrix of the SM extended with three singlet neutrinos. After electroweak
symmetry breaking, fermion mass terms, among all fermionic fields in
the two-spinor language, are in four separated groups --- the up- and
down-type quarks, charged leptons, and neutrinos. Each group has
a (two-)spinor mass matrix with six states, three coming from 
$SU(2)_{L}$ doublets while another three from singlets. That
is from the mysterious fact about having three families of SM
fermions while each family is given simply as the minimal chiral set 
with all anomalies canceled nontrivially among the fermions \cite{GM,O}.\footnote{This important feature of the SM is less than
appreciated wide enough. For instance, a lecture series available just
a few years ago asked the question and  gave a presentation that
the fermion in a SM family fails by a little to meet the criterion,
without any reference to the earlier work establishing the notion.}
It is natural to put each of the $6\times 6$ mass matrices in the
form with $3\times 3$ doublet and singlet blocks as
\beq
 {\mathcal M}
=
\left( \begin{array}{cc} 
{\mathcal M}_{\!\ssc L} &{{\mathcal M}_{\!\ssc D}} \\
{{\mathcal M}_{\!\ssc D}} & {\mathcal M}_{\!\ssc S} 
 \end{array}\right) \;,
\eeq
where ${\mathcal M}_{\!\ssc L}$ and ${\mathcal M}_{\!\ssc S}$
denote the doublet and the singlet block while ${\mathcal M}_{\!\ssc D}$
is the doublet-singlet mixing block. For the quark and charged lepton 
sectors, ${\mathcal M}_{\!\ssc L}$ and ${\mathcal M}_{\!\ssc S}$ have to
be vanishing to maintain electric charge and color conservation. The 
$3\times 3$ block  ${\mathcal M}_{\!\ssc D}$ is the 
`Dirac mass matrix' usually discussed. For the neutrino sector,
however, both ${\mathcal M}_{\!\ssc L}$ and ${\mathcal M}_{\!\ssc S}$
may be nonzero. ${\mathcal M}_{\!\ssc L}$ with mass terms
involving SM neutrinos ($\nu_e$, $\nu_\mu$, and 
$\nu_\tau$)
only may get nonzero contributions from dimension-five term involving
two Higgs bosons and two leptonic doublets which can be obtained from integrating
out a heavy scalar triplet for example.
\footnote{The term `flavor eigenstates' or `(weak-)interaction 
eigenstates' is used to called the three states
quite often in the recent literature. The terminology is also very 
misleading. There is nothing `eigen' about a `flavor' state. The only 
set of independent states distinguishing themselves out for arbitrary linear 
combinations for the three SM neutrino states, or say charged lepton
states for that matter, are the mass eigenstates. $\nu_e$, $\nu_\mu$, 
and $\nu_\tau$ are of course distinguished as exact $SU(2)_{L}$
partners of the charged lepton mass eigenstates, and nothing more than
that. We really think the term `flavor eigenstates' should not be used
at all within the SM. In models with extra family/flavor symmetries,
a particular linear combination may of course has its specific 
family/flavor quantum number.} The naive notion of Dirac neutrinos
is from the case of vanishing ${\mathcal M}_{\!\ssc L}$ and 
${\mathcal M}_{\!\ssc S}$. That the idea of calling a SM neutrino (part of)
a Dirac particle only for the case is beyond the simple definition of
what is a Dirac fermion in quantum field theory, as discussed above.
 
\section{Conclusion}
The key point that motivated this article can be summarized by the simple 
mathematics as given in a single equation, Eq.(\ref{key}). It presents the 
statement of the mathematical equivalence of the mass term, in the usual 
textbook four-spinor form, for two degenerate Majorana fermions and that of 
one Dirac fermion. To establish that beyond any doubt for those seeing that 
as completely surprising, and review the related issues of mass terms for 
spinor fields, we write the full article. One can see that there are many 
subtleties involved, and that quite some ambiguities linger around the related 
terminology. All these easily lead to confusions among students or even
some professional physicists. 

The SM quark and charged lepton mass eigenstates can be described by 
Dirac fermions the left- and right-handed parts of which come from
$SU(2)_{L}$ doublets and singlets, respectively. To write 
any such fermion as a pair of Majorana ones, the mass eigenstates 
would be the linear combinations of parts with conjugate charges, hence
undesirable. For the case of the neutrinos, or fermion masses in
general, the Majorana versus Dirac description could sound arbitrary.
A quite common practice among phenomenology papers
is to call neutrino masses Dirac if each of the mass terms involve 
neutrino states from an $SU(2)_{L}$ doublet and a singlet, and
Majorana if there is a mass term involving only $SU(2)_{L}$ doublet 
states. The unstated rule, while is consistent and matches well with
the other fermionic sector, does not have a firm basis in generic
quantum field theory, and cannot be unambiguously extended when
a third kind of states is involved. 

Generic fermion mass matrices are to be described, as in the case of 
the SM, in terms of two-spinor fields. The `Dirac' or `Majorana' mass
matrices of which are not mass matrices for the corresponding 
four-spinors. The latter can only be formally constructed in terms of
two-spinor `exact' mass eigenstates, defined as normalized states 
giving the mass matrices as diagonal with real and positive 
(eigenvalue) entries. The eigenstates are exact up to the phase factors.
Linear combinations of such `exact' mass eigenstates of a degenerate 
eigenvalue are not eigenstates of the same eigenvalue. The basic mass
matrix $\mathcal{M}$ is not hermitian. It does not represent a physical 
observable. Only $\mathcal{M}^\dag\mathcal{M}$ and 
$\mathcal{M}\mathcal{M}^\dag$ represent the observable of mass-squared. 
The latter are representations of the same operator
which is the relevant Casimir operator of the Poincar\'e symmetry.

\section{Acknowledgments}
The authors are partially supported by research grants numbered 
NSC 99-2112-M-008-003-MY3 and NSC 102-2112-M-008-007-MY3 of the 
National Science Council of Taiwan. Y.C. is further supported by 
grant numbered   NSC 103-2811-M-008-018 from the same agency.
We thank T.-M.~Yan for questions which motivated us to elaborate
the matter in full detail as presented here. O.K. thanks also the
organizers of the 2013 Cross-Strait Meeting on Particle Physics and
Cosmology in an invited talk of which a major part of the results
here have been presented, and A. Das for his comments and suggestions
on the manuscript.



\begin{thebibliography}{99}
\bibitem{WB}
J. Wess, J. Bagger, Supersymmetry and Supergravity, Princeton University Press, 1992; the notation for the metric and the gamma matrices in the text 
differs from that by Maggiore, but the $\sigma^i=-\bar{\sigma}^i$ are the same and all 
details of the two-spinor notation compatible.
\end{thebibliography}


\end{document}